\documentclass[prd,aps,floats,floatfix,nofootinbib]{revtex4}
\usepackage{amsmath}
\usepackage{amsfonts}
\usepackage{graphicx}
\usepackage{graphicx}
\usepackage{dcolumn}
\usepackage{bm}
\usepackage[dvips]{color}
\usepackage{slashed}
\newcommand{\beq}{\begin{equation}}
\newcommand{\eeq}{\end{equation}}
\begin{document}
%%%%%%%%%%%%%%%%%
\newcommand{\vect}[1]{\overrightarrow{#1}}
\newcommand{\smbox}[1]{\mbox{\scriptsize #1}}
\newcommand{\tanbox}[1]{\mbox{\tiny #1}}
\newcommand{\vev}[1]{\langle #1 \rangle}
\newcommand{\Tr}[1]{\mbox{Tr}\left[#1\right]}
\newcommand{\cosb}{c_{\beta}}
\newcommand{\sinb}{s_{\beta}}
\newcommand{\tanb}{t_{\beta}}
\newcommand{\picwidth}{3.4in}
%%%%%%%%%%%%%%%%

\preprint{MSU-HEP-080508}
\title{The Limits of Custodial Symmetry}

\author{R. Sekhar Chivukula$^{a,b}$}
\email[]{sekhar@msu.edu}
\author{Stefano Di Chiara$^{a,c}$}
\email[]{dichiara@cp3.sdu.dk}
\author{Roshan Foadi$^a$}
\email[]{foadiros@msu.edu}
\author{Elizabeth H. Simmons$^{a,b}$}
\email[]{esimmons@msu.edu}
\affiliation{$^a$Department of Physics and Astronomy,
Michigan State University, East Lansing, MI 48824, USA\\
$^b$School of Natural Sciences, Institute for Advanced Study, Princeton, NJ 08540 USA\\
$^c$ CP3-Origins, Campusvej 55, DK-5230 Odense M, Denmark.
}
\date{October 13, 2009}
%\date{\today}

\begin{abstract}
We introduce a toy model implementing the proposal of using a custodial symmetry to protect the $Z b_L \bar{b}_L$ coupling from large corrections.  This ``doublet-extended standard model" adds a weak doublet of fermions (including a heavy partner of the top quark) to the particle content of the standard model in order to implement an $O(4) \times U(1)_X  \sim SU(2)_L \times SU(2)_R \times P_{LR} \times U(1)_X$ symmetry in the top-quark mass generating sector.  This symmetry is softly broken to the gauged $SU(2)_L \times U(1)_Y$ electroweak symmetry by a Dirac mass $M$ for the new doublet; adjusting the value of $M$ allows us to explore the range of possibilities between the $O(4)$-symmetric ($M \to 0$) and standard-model-like ($M \to \infty$) limits.  In this simple model, we find that the experimental limits on the $Z b_L \bar{b}_L$ coupling favor smaller $M$ while the presence of a potentially sizable negative contribution to $\alpha T$  strongly favors large $M$.  Comparison with precision electroweak data shows that the heavy partner of the top quark must be heavier than about 3.4 TeV, making it difficult to search for at LHC. This result demonstrates 
that electroweak data strongly limits the
amount by which the custodial symmetry of the top-quark mass generating sector can be enhanced relative
to the standard model. Using an effective field theory calculation, we illustrate how the leading contributions to $\alpha T$, $\alpha S$ and the $Zb_L \bar{b}_L$ coupling in this model arise from an effective operator coupling right-handed top-quarks to the
$Z$-boson, and how the effects on these observables are correlated.
We contrast this toy model with extra-dimensional models in which
the extended custodial symmetry is invoked to control the size of additional contributions to $\alpha T$ and
the $Zb_L\bar{b}_L$ coupling, while leaving the standard model contributions essentially unchanged.
\end{abstract}

\maketitle
\section{Introduction}
Although the standard model (SM) is in excellent agreement with the experimental data, the triviality and naturalness problems in the Higgs sector demonstrate that it is, at best, an effective field theory valid up to some energy scale $\Lambda$.  The literature contains a rich variety of ideas about what kind of new physics might subsume or augment the SM at energies above those explored by current experiments.  In building models of new physics, incorporating the large mass of the top quark while still conforming to the precision electroweak data remains a challenge.  In particular, the interactions introduced to give rise to the top quark mass
typically introduce corrections to the $Zb_L\bar{b}_L$ coupling, $g_{Lb}$.  Not only is $g_{Lb}$ tightly constrained by experiment, but the value predicted at the one-loop level in the SM is already about $2\sigma$ away from the central experimental value -- so that radiative corrections of the wrong sign will tend to push the theoretical value further from agreement with experiment.

Agashe et al. \cite{Agashe:2006at} have shown that the constraints on beyond the standard model physics related to the $Zb_L\bar{b}_L$ coupling can, in principle, be loosened if the global $SU(2)_L \times SU(2)_R$ symmetry of the electroweak symmetry breaking sector is actually a subgroup of a larger global symmetry  of both the symmetry breaking and top quark mass generating sectors of the theory. In particular, they propose that these interactions preserve an  $O(4)\sim SU(2)_L\times SU(2)_R\times P_{LR}$ symmetry, where $P_{LR}$ is a parity interchanging $L\leftrightarrow R$. The $O(4)$ symmetry is then spontaneously broken to  $O(3)\sim SU(2)_V\times P_{LR}$,  breaking the elecroweak interactions but
protecting $g_{Lb}$ from radiative corrections, so long as the left-handed bottom quark is a $P_{LR}$ eigenstate.

In this paper we construct an explicit realization of the simplest $O(4)$-symmetric extension of the SM.  For reasons that will shortly become clear, we call this model the doublet-extended standard model or DESM.  Because the DESM is minimal, it displays the essential ingredients protecting $g_{Lb}$ without the burden of additional states, interactions, or symmetry patterns that might otherwise obscure the role played by custodial $O(3)$.  Because it is concrete, it also enables us to explore how the new symmetry impacts the model's ability to conform with the constraints imposed by other precision electroweak data.

In our model, all operators of dimension-4 in the Higgs potential and the sector generating the top quark mass respect a global $O(4)\times U(1)_X$ symmetry; the $U(1)_X$ enables the SM-like fermions to obtain the appropriate electric charges and hypercharges.  In addition to the particle content of the SM, we introduce a new weak doublet of Dirac fermions, $\Psi=(\Theta,T^\prime)$, and combine $\Psi_L$  with the left-handed top-bottom doublet $(t^\prime_L, b_L)$ to form a $(2,2^*)$ under the global $SU(2)_L\times SU(2)_R$ symmetry.  The $b_L$ state is thereby endowed with identical charges under the two global $SU(2)$ groups, $T_L^3=T_R^3$, making it a parity eigenstate, as desired.   We also find that the $T'$ mixes with $t^\prime$ to form a SM-like top quark and a heavy partner.  The $O(4) \times U(1)_X$ symmetric Yukawa interaction can, of course, be extended to the bottom quark and the remaining electroweak doublets, by adding further spectator fermions;  here we focus exclusively on the partners of the top quark since they give the dominant contribution to $g_{Lb}$.

To enable electroweak symmetry breaking and fermion mass generation to proceed, the global symmetry is explicitly broken to $SU(2)_L\times U(1)_Y$ by a dimension-three Dirac mass $M$ for $\Psi$. As $M\to\infty$ the ordinary SM top Yukawa interaction is recovered; as $M\to 0$ the model becomes exactly $O(4)\times U(1)_X$ symmetric; adjusting the value of $M$ allows us to interpolate between these extremes and to investigate the limits to which
the custodial symmetry of the top-quark mass generating sector can be enhanced.  
When we calculate the dominant one-loop corrections to $g_{Lb}$ in our model, we find, consistent with ref.  \cite{Agashe:2006at}, that because $b_L$ is a $P_{LR}$ eigenstate,  $g_{Lb}$ is protected from radiative corrections in the $M\to 0$ limit and these corrections return as $M$ is switched on.   However, when we study the behavior of oblique radiative corrections as $M$ is varied, we find that in the small-$M$ limit where $g_{Lb}$ is closer to the experimental value, the oblique corrections become unacceptably large. In particular, in the
$M \to 0$ limit the enhanced custodial symmetry produces a potentially sizable negative contribution to $\alpha T$.  Using effective field theory methods, we illustrate how the leading contributions to $\alpha T$, $\alpha S$ and the $Zb_L \bar{b}_L$ coupling in this model 
arise from a single effective operator coupling right-handed top quarks to the $Z$ boson. We then
contrast this toy model with extra-dimensional models in which
the extended custodial symmetry is invoked to control the size of additional contributions to $\alpha T$ and
the $Zb_L\bar{b}_L$ coupling, while leaving the standard model contributions essentially unchanged.

The paper is organized as follows: In Sec.~(\ref{sec:DESM}) we review the concept of using custodial symmetry to protect $g_{Lb}$ from large corrections, and then present the doublet extended standard model (DESM) as a concrete realization of this idea. In Sec.~(\ref{sec:pheno}) we calculate $g_{Lb}$ at one loop order, and compare it with the experimentally measured value $g_{Lb}^{ex}$; indeed, varying $M$ to move away from the SM limit does allow $g_{Lb}$ to approach the experimental value more closely.  In Sec.~(\ref{sec:pheno}) we calculate the oblique electroweak parameters $\alpha S$ and $\alpha T$, finding that the latter provides tighter constraints on $M$ that push the model back towards the $SM$ limit.   We then compare the DESM's joint prediction for $\alpha S$ and $\alpha T$ to the region of the $\alpha S - \alpha T$ plane that gives the best fit to existing data \cite{Amsler:2008zzb} and find that the DESM is most consistent with experiment in the limit where it most closely approximates the SM. Using effective field
theory, in Sec.~(\ref{sec:Effective}) we compute the leading-log contributions to 
$\alpha T$, $\alpha S$ and the $Zb_L\bar{b}_L$ coupling in the DESM.  We demonstrate that 
 these corrections all arise from a single effective operator coupling right-handed top-quarks to the
$Z$-boson and are therefore correlated.   In Sec. (\ref{appxB}) we contrast the DESM  toy model with extra-dimensional models in which the extended custodial symmetry is invoked to control the size of additional contributions to $\alpha T$ and
the $Zb_L\bar{b}_L$ coupling,
relating our results to those previously discussed in  \cite{Carena:2006bn,Carena:2007ua} and
commenting on the  model-independent analysis presented in ref.~\cite{Pomarol:2008bh}.  Sec.~(\ref{sec:Conclusion5})  summarizes our results and presents our conclusions. 

\section{Doublet-Extended Standard Model}
\label{sec:DESM}
\subsection{Custodial Symmetry and Z coupling}\label{cus}

The tree-level coupling of a SM fermion $\psi$ to the $Z$ boson is, 
\begin{equation}
\frac{e}{c_w s_w}\ (T^3_L-Q \sin^2\theta_W) \ Z^\mu \bar{\psi}\gamma_\mu \psi \ ,\ \ 
\label{eq:coupling}
\end{equation}
where $T^3_L$ and $Q$ are, respectively, the weak isospin and electromagnetic charges of fermion $\psi$, $e$ is the electromagnetic coupling; $c_w$ and $s_w$ are the cosine and sine of the weak mixing angle.  Because the electromagnetic charge is conserved, loop corrections to the $Z\bar{\psi}\psi$ coupling do not alter it; however, the weak symmetry $SU(2)_L$ is broken at low energies, and radiative corrections to the $T^3_L$ coupling are present in the SM.

Following the proposal of \cite{Agashe:2006at}, we wish to construct a scenario in which the $T^3_L$ coupling is not subject to flavor-dependent radiative corrections.  To start, we note that the accidental custodial symmetry of the SM implies that the vectorial charge $T^3_V\equiv T^3_L+T^3_R$ is conserved 
\begin{equation}
\delta T^3_V=\delta T^3_L+\delta T^3_R=0 \ .
\label{eq:deltaT3v}
\end{equation}
This suggests a way to evade flavor-dependent corrections to $T_L^3$ itself, by adding a parity symmetry $P_{LR}$ that exchanges  $L\leftrightarrow R$.  If $\psi$ is an eigenstate of this parity symmetry and the symmetry persists at the energies of interest, then
\begin{equation}
\delta T^3_L=\delta T^3_R  \ .
\end{equation}
Now, we see that Eq. (\ref{eq:deltaT3v}) is satisfied by having the two terms on the RHS vanish separately, rather than remaining non-zero and canceling one another.  In other words, $\delta T^3_L = 0$ and the $Z\bar{\psi}\psi$ coupling remains fixed even to higher-order in this scenario.   We will now show how to implement this idea for the $b$-quark in a toy model and examine the phenomenological consequences.

\subsection{The Model}

Let us construct a simple extension of the SM that implements this parity idea for the third-generation quarks, in order to suppress radiative corrections to the $Zb\bar{b}$ vertex.  We extend the global $SU(2)_L \times SU(2)_R$ symmetry of the Higgs sector of the SM to an $O(4)\times U(1)_X\sim SU(2)_L\times SU(2)_R\times P_{LR} \times U(1)_X$ for both the symmetry breaking and top quark mass generating sectors of the theory.  As usual, only the electroweak subgroup, $SU(2)_L\times U(1)_Y$, of this global symmetry is gauged; our model does not include additional electroweak gauge bosons.  The global $O(4)$ spontaneously breaks to $O(3) \sim SU(2)_V \times P_{LR}$ which will protect $g_{Lb}$ from radiative corrections, as above, provided that the left-handed bottom quark is a parity eigenstate:  $P_{LR} b_L = \pm b_L$.  The additional global $U(1)_X$ group is included to ensure that the light $t$ and $b$ eigenstates, the ordinary top and bottom quarks, obtain the correct hypercharges.

In light of the extended symmetry group, the relationships between electromagnetic charge $Q$, hypercharge $Y$, the left- and right-handed $T^3$ charges, and the new charge $Q_X$ associated with $U(1)_X$ are as follows:
\begin{eqnarray}
Y & = & T^3_R+Q_X \ , \label{eq:hyper} \\
Q & = & T^3_L+Y = T^3_L+T^3_R+Q_X \ . \label{eq:EM}
\end{eqnarray}
Since the $b_L$ state is supposed to correspond to the familiar bottom-quark, it has the familiar SM charges $T^3_L(b_L) = - 1/2$, and $Q(b_L) = -1/3$, and $Y(b_L) = 1/6$. Because $b_L$ must be an eigenstate under $P_{LR}$, we deduce that $T^3_R(b_L) = T^3_L(b_L) = -1/2$.  Then to be consistent with Eqs. (\ref{eq:hyper}) and (\ref{eq:EM}), its charge under the new global $U(1)_X$ must be $Q_X(b_L) = 2/3$.   Moreover, since the left-handed $b$ quark is an $SU(2)_L$ partner of the left-handed $t$ quark,  the full left-handed top-bottom doublet must have the charges $T^3_R = -1/2$ and $Q_X = 2/3$, just as the full doublet has hypercharge $Y = 1/6$.  Finally, the non-zero $T_3^R$ charge of the top-bottom doublet tells us that this doublet forms part of a larger multiplet under the $SU(2)_L \times SU(2)_R$ symmetry and it will be necessary to introduce some new fermions with $T^3_R = 1/2$ to complete the multiplet.

We therefore introduce a new doublet of fermions $\Psi  \equiv (\Omega, T^\prime)$.  The left-handed component, $\Psi_L$ joins with the top-bottom doublet $q_L \equiv (t_L^\prime, b_L)$ to form an $O(4)\times U(1)_X$ multiplet
\begin{equation}
  {\cal Q}_L = \left( {\begin{array}{*{20}c}
   {t^\prime_L } & {\Omega_L }  \\
   {b_L } & {T^\prime_L }  \\
 \end{array} } \right)
\equiv \left(\begin{array}{cc} q_L & \Psi_L \end{array}\right) \ ,
\end{equation}
which transforms as a $(2,2^*)_{2/3}$ under $SU(2)_L\times SU(2)_R\times U(1)_X$.  The parity operation $P_{LR}$, which exchanges the $SU(2)_L$ and $SU(2)_R$
transformation properties of the fields, acts on ${\cal Q}_L$ as:
\begin{equation}
P_{LR} {\cal Q}_L = - \left[ ( i \sigma_2)\, {\cal Q}_L\, (i \sigma_2) \right]^T = \left( {\begin{array}{*{20}c}
   {T^\prime_L } & {-\Omega_L }  \\
   {-b_L } & {t^\prime_L }  \\
 \end{array} } \right)
\end{equation}
exchanging the diagonal components, while reversing the signs of the off-diagonal components. Thus $t^\prime_L$ and $T^\prime_L$ are constrained to share the same electromagnetic charge, in order to satisfy Eq.~(\ref{eq:EM}).  In fact, we will later see that the $t^\prime$ and $T^\prime$ states mix to form mass eigenstates corresponding to the top quark ($t$) and a heavy partner ($T$).  The charges of the components of ${\cal Q}_L$ are listed in Table \ref{tab:charges}.

\begin{table}[bt]
\begin{tabular}{|c||c|c|c|c||c|c|c|c|}
\hline
&$t^\prime_L$&$b_L$&$\Omega_L$ & $T^\prime_L$&$t^\prime_R$&$b_R$&$\Omega_R$ & $T^\prime_R$\\
\hline
$T^3_L$& $\frac12$ & $-\frac12$ & $\frac12$ & $-\frac12$ & $0$ & $0$ & $\frac12$ & $-\frac12$  \\
\hline
$T^3_R$& $-\frac12$ & $-\frac12$ & $\frac12$ & $\frac12$ & $0$ & $-1$ & $0 $ & $0$\\
\hline
$Q$& $\frac23$ & $-\frac13$ & $\frac53$ &$\frac23$ &$\frac23$ & $-\frac13$ & $\frac53$ &$\frac23$ \\
\hline
$Y$& $\frac16$ & $\frac16$ & $\frac76$ & $\frac76$ & $\frac23$ & $-\frac13$ & $\frac76$ &$\frac76$\\
\hline
$Q_X$& $\frac23$ &$\frac23$ &$\frac23$ &$\frac23$ &$\frac23$ &$\frac23$ &$\frac76$ &$\frac76$ \\
\hline
\end{tabular}
\caption{\label{tab:charges} Charges of the fermions under the various symmetry groups in the model. Note that, as discussed in the text, other $T^3_R$ and $Q_X$ assignments for the $\Omega_R$ and $T^\prime_R$ states are possible.}
\end{table}

We assign the minimal right-handed fermions charges that accord with the symmetry-breaking pattern we envision:  the top and bottom quarks will receive mass via Yukawa terms that respect the full $O(4) \times U(1)_X$ symmetry, while the exotic states will have a dimension-three mass term that explicitly breaks the large symmetry to $SU(2)_L \times U(1)$.   Moreover, to accord with experiment, the  $t'_R$ and $b_R$ must have $T^3_L = 0$ and share the electric charges of their left-handed counterparts.  The top and bottom quarks will receive mass through a Yukawa interaction with a SM-like Higgs multiplet that breaks the electroweak symmetry. The simplest choice is to assign the Higgs multiplet to be neutral under $U(1)_X$; in this case, both $t'_R$ and $b_R$ share the $Q_X = 2/3$ charge of $t'_L$ and $b_L$.  Therefore, from equations Eq. (\ref{eq:hyper}) and (\ref{eq:EM}), we find $T^3_R(t'_R) = 0$ (meaning that $t'_R$ can be
chosen to be an $SU(2)_R$ singlet) and $T^3_R(b_R)$ = -1 (so that $b_R$ is, minimally, part of an $SU(2)_R$ triplet if we extend the symmetry to the bottom quark mass generating sector).  Turning now to the $T'_R$ and $\Omega_R$ states, we see that they must form an $SU(2)_L$ doublet with hypercharge $7/6$ so that the Dirac mass term for $\Psi$ preserves the electroweak symmetry as desired.\footnote{This means that the $\Omega_R$ and $T^\prime_R$ states do not fill out the $SU(2)_R$ triplet to which $b_R$ belongs -- which is uncharged under $SU(2)_L$ and carries hypercharge $2/3$; other exotic fermions must play that role if we wish to extend the symmetry to the bottom quark mass generating sector.}  Finally, we choose $T^3_R(\Omega_R) = T^3_R(T^\prime_R) = 0$, which implies $Q_X = 7/6$ for both states, as the minimal choice satisfying the constraint imposed by Eq. (\ref{eq:hyper}); other choices of $T^3_R$ charge would involve adding additional fermions to form complete $SU(2)_R$ multiplets.  The charges of the fermions are listed in Table \ref{tab:charges}.

Now, let us describe the symmetry-breaking pattern and fermion mass terms explicitly.  
Spontaneous electroweak symmetry breaking proceeds through a Higgs multiplet that transforms as a $(2,2^*)_0$ under $SU(2)_L \times SU(2)_R \times U(1)_X$:
\begin{equation}
 \Phi  = \frac{1}{\sqrt{2}}\left( {\begin{array}{*{20}c}
   {v+h + i \phi^0 } & {i\sqrt{2}\ \phi^+  }  \\
   {i\sqrt{2}\ \phi^-  } & {v+h-i\phi^{0} }  \\
 \end{array} } \right)\ .
 \label{eq:higgsdef}
\end{equation}
Again, the parity operator $P_{LR}$ exchanges the diagonal fields and reverses the signs of the off-diagonal elements.  When the Higgs acquires a vacuum expectation value, the longitudinal  $W$ and $Z$ bosons acquire mass and a single Higgs boson remains in the low-energy spectrum.
The Higgs multiplet has an $O(4)\times U(1)_X$ symmetric Yukawa interaction with the top quark:
\begin{equation}
{\cal{L}}_{\rm Yukawa}= - \lambda_t \text{Tr} \left( \overline{\cal Q}_L\cdot \Phi\right) t^\prime_R \ + \rm{h.c.}\ .
\label{eq:Yuk}
\end{equation}
that contributes to generating a top quark mass. In principle, the same Higgs multiplet can also contribute to the bottom quark mass through a separate, and similarly $O(4)\times U(1)_X$ symmetric, Yukawa interaction involving the $SU(2)_R$ triplet to which $b_R$ belongs.  Since the phenomenological issues that concern us in this paper are affected far more strongly by $m_t$ than by the far-smaller $m_b$, we will neglect this and any other Yukawa interaction.

Next we break the  full $O(4)\times U(1)_X$ symmetry to its electroweak subgroup. We do so
first by gauging $SU(2)_L\times U(1)_Y$.  In addition, we wish to preserve the $O(4)$ symmetry
of the top quark mass generating sector in all dimension-4 terms, but break it 
softly by introducing a dimension-3 Dirac mass term for  $\Psi$,
\begin{equation}
{\cal{L}}_{\rm mass}= - M\ \bar{\Psi}_L\cdot\Psi_R + h.c. 
\label{eq:M}
\end{equation}
that explicitly breaks the global symmetry to $SU(2)_L\times U(1)_Y$.  Note that we therefore expect that any flavor-dependent radiative corrections to the $Zb_L\bar{b}_L$ coupling will vanish in the limit $M \to 0$, as the protective parity symmetry is restored; alternatively, as $M \to \infty$, the larger symmetry is pushed off to such high energies that the resulting theory looks more and more like the SM.

In addition to the fermions explicitly described above, a more complete version of this toy model must contain several other fermions to fill out the $SU(2)_R$ multiplet to which the $b_R$ belongs and also some spectator fermions that cancel $U(1)$ anomalies.  However, the  toy model suffices for exploration of the issues related to the $Zb_L\bar{b}_L$ coupling that is the focus of this paper.

\subsection{Mass Matrices and Eigenstates}

When the Higgs multiplet acquires a vacuum expectation value and breaks the electroweak symmetry, masses are generated for the top quark, its heavy partner $T$ and the exotic fermion $\Omega$ through the mass matrix:
\begin{equation}
{\cal L}_{\rm mass} = -\left(\begin{array}{*{20}c} t^\prime_L & T^\prime_L \end{array}\right)\ 
\left({\begin{array}{*{20}c} m & 0  \\ m & M \\ \end{array} } \right)
\left(\begin{array}{c} t^\prime_R \\ T^\prime_R \end{array}\right) -M \bar{\Omega}_L \Omega_R + {\rm h.c} \ ,
\end{equation} 
where
\begin{equation}
m = \frac{{\lambda _t v}}{{\sqrt 2 }} \ .
\label{eq:mtSM}
\end{equation} 
Note that the  $\Omega$ field is decoupled from the SM sector, and its mass is simply $m_\Omega = M$.
The bottom quark remains massless because we have ignored its Yukawa coupling.

Diagonalizing the top quark mass matrix yields mass eigenstates $t$ (corresponding to the SM top quark) and $T$ (a heavy partner quark), with corresponding eigenvalues
\begin{equation}
m_t^2 = \frac{1}{2}\left[1-\sqrt{1+\frac{4m^4}{M^4}}\right]M^2+m^2 \ , \ \ \ \ \ \ \ \ 
m_T^2 = \frac{1}{2}\left[1+\sqrt{1+\frac{4m^4}{M^4}}\right]M^2+m^2 \ .
\label{eq:massEv}
\end{equation}
The mass eigenstates are related to the original gauge eigenstates through the rotations:
\begin{equation}
\left(\begin{array}{c} t^\prime_R \\ T^\prime_R \end{array}\right) =
\left(\begin{array}{cc} \cos\theta_R & \sin\theta_R \\ -\sin\theta_R & \cos\theta_R \end{array}\right)
\left(\begin{array}{c} t_R \\ T_R \end{array}\right) \ ,\ \ \ \ 
\left(\begin{array}{c} t^\prime_L \\ T^\prime_L \end{array}\right) =
\left(\begin{array}{cc} \cos\theta_L & \sin\theta_L \\ -\sin\theta_L & \cos\theta_L \end{array}\right)
\left(\begin{array}{c} t_L \\ T_L \end{array}\right) \ ,
\label{eq:massEs}
\end{equation}
whose mixing angles are given by 
\begin{equation}
\sin\theta_R=\frac{1}{\sqrt{2}}\sqrt{1-\frac{1-2m^2/M^2}{\sqrt{1+4m^4/M^4}}} \ ,\ \
\sin\theta_L=\frac{1}{\sqrt{2}}\sqrt{1-\frac{1}{\sqrt{1+4m^4/M^4}}} \ .
\end{equation}
From these equations the decoupling limit $M\to\infty$ is evident: $m_t$ approaches its SM value as in Eq.~(\ref{eq:mtSM}), the $t-T$ mixing goes to zero, and $T$ becomes degenerate with $\Omega$. Conversely, in the limit $M\to 0$, the full $O(4)\times U(1)_X$ symmetry is restored and only the combination $T'_L + t'_L$ couples to $t_R$ with mass $m$.

\begin{figure}
\begin{center}
\includegraphics[width=5in]{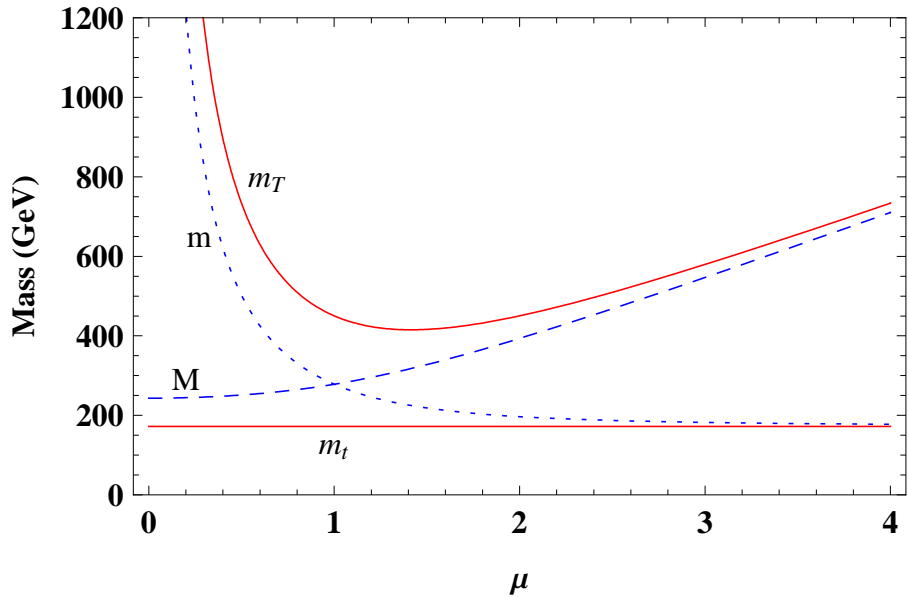}
\end{center}
\caption{The curves show the behaviors of $m$ (dotted), $M$ (dashed), and $m_T$ (upper solid) as functions of $\mu\equiv M/m$ when $m_t$ is held fixed.  The solid horizontal line corresponds to $m_t\simeq$ 172 GeV.}
\label{fig:masses}
\end{figure}

For phenomenological discussion, it will be convenient to fix $m_t$ at its experimental value and express the other masses in terms of $m_t$ and the ratio $\mu\equiv M/m$. Fig.~\ref{fig:masses} shows how $m$, $M$, and $m_T$, vary with $\mu$; the horizontal line represents $m_t$ which is being held fixed at 172 GeV.  In the limit as $\mu$ becomes large, $m\to m_t$, $m_T \sim M$ grows steadily, and the mixing angles decline toward zero; this is a physically-sensible limit that ultimately leads back to the SM.  However we see that the opposite limit, where $\mu\to 0$ can only be achieved for $m\to\infty$, which is not physically reasonable since it corresponds to taking $\lambda_t \to \infty$.  Hence, we will need to take care in talking about the case of small $\mu$.

\section{Phenomenology}
\label{sec:pheno}
\subsection{$Z$ coupling to $b_L \bar{b}_L$}
\label{sec:gLb}

We are now ready to study how the flavor-dependent corrections to the $Zb_L \bar{b}_L$ coupling behave in our toy model.  Specifically, if we write the  $Zb_L\bar{b}_L$ coupling as
\begin{equation}
\frac{e}{c_w s_w} \ \left(- \frac{1}{2} + \delta g_{Lb} + \frac{1}{3}\sin ^2 \theta _L \right)\ Z_\mu\  \bar{b}_L\gamma_\mu b_L \ ,
\label{eq:totalZbb}
\end{equation}
then all the flavor-dependence is captured by $\delta g_{Lb}$.  At tree-level, the $Z b_L \bar{b}_L$ coupling in our model has its SM value, with $\delta g_{Lb} = 0$, because the $b_L$ has the same quantum numbers as in the SM.  However, at one-loop, flavor-dependent vertex corrections arise and these give non-zero corrections to $\delta g_{Lb}$; these corrections differ from those in the SM due to the presence of vertex corrections involving exchange of $T$, the heavy partner of the top quark.

\begin{figure}
\begin{center}
\includegraphics[width=2.2 in]{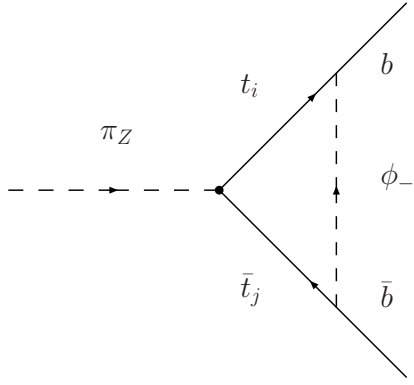}
\end{center}
\caption{One-loop vertex correction diagram for $\pi_Z \rightarrow b \bar{b}$ in our model.  The $t_{i,j}$ may be either the top quark ($t$) or its heavy partner ($T$).}
\label{fig:chibb}
\end{figure}

The calculation may be done conveniently in the ``gaugeless'' limit \cite{Barbieri:1992nz,Barbieri:1992dq,Oliver:2002up,Abe:2009ni}, in which the $Z$ boson is treated as a non-propagating external field coupled to the current $j^\mu_{3L} - j^\mu_Q \sin^2 \theta_L$.  Operationally, this involves replacing $Z_\mu$ with $\partial_\mu\phi^0/m_Z$ in the gauge current interaction, where $\phi^0$ is the Goldstone boson eaten by the $Z$:
\begin{equation}
\frac{e}{c_w s_w}\ Z_\mu (j^\mu_{3L} - j^\mu_Q \sin^2 \theta_L)\ \ \to\ \  \frac{e}{c_w s_w m_Z}\ \partial_\mu\phi^0 (j^\mu_{3L} - j^\mu_Q \sin^2 \theta_L)
= \frac{2}{v}\ \partial_\mu\phi^0 (j^\mu_{3L} - j^\mu_Q \sin^2 \theta_L)
\label{eq:gaugeless}
\end{equation} 
The general vertex diagram shown in Fig.~\ref{fig:chibb}, will yield radiative corrections to the effective operator  $\partial_\mu\phi^0\ \bar{b}_L\gamma^\mu b_L$; that is, the expression for this diagram will include a term of the form
\beq
A\ \partial_\mu\phi^0\ \bar{b}_L\gamma^\mu b_L \ .
\label{api}
\eeq
Comparing the last three equations shows that the coefficient $A$ is proportional to the quantity we are interested in: 
\beq
\delta g_{Lb}=\frac{v}{2}\ A \ .
\eeq

\begin{figure}[b]
\begin{center}
\includegraphics[width=4 in]{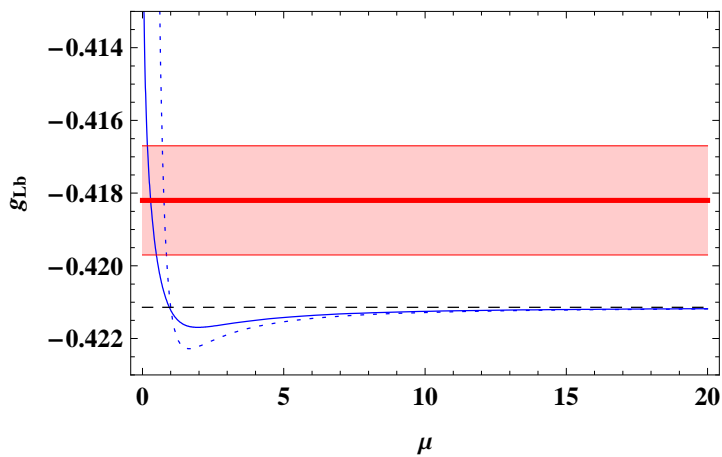}
\end{center}
\caption{The solid (blue) curve shows the DESM model's prediction for $g_{Lb}$, Eq.~(\ref{eq:dglb}). The thick horizontal line corresponds to $g_{Lb}^{ex}=-0.4182$, while the two horizontal upper and lower solid lines bordering the shaded band correspond to the $\pm 1\sigma$ deviations \cite{:2005ema}. The SM prediction is given by the dashed horizontal line. The leading-log contribution, Eq. (\ref{eq:zbbleading}), is shown by the dotted curve.}
\label{fig:gZbb}
\end{figure}

We have calculated the several loop diagrams represented by Fig.~\ref{fig:chibb} and obtain the following expression for $\delta g_{Lb}$:
\begin{eqnarray}
\delta g_{Lb} & = & \frac{m_t^2}{16\pi^2 v^2}\left[\cos^2\theta_L\left(\cos 2\theta_L+\sin^2\theta_R\right)
+\frac{m_T^2}{m_t^2}\sin^2\theta_L \left(\cos^2\theta_R-\cos 2\theta_L\right) \right. \nonumber \\
& &\quad\quad\quad\quad -\left. \frac{m_T/m_t}{2}\sin 2\theta_L\left(\frac{m_T^2/m_t^2+1}{2}\sin 2\theta_R-2\frac{m_T}{m_t}\sin 2\theta_L\right)
\frac{\log(m_T^2/m_t^2)}{m_T^2/m_t^2-1}\right]\,,
\label{eq:dgZbb}
\end{eqnarray} 
where the prefactor proportional to $m_t^2$ is the SM result for this class of diagram.  We expect to see  $\delta g_{Lb}$ vanish in the limit $M \to 0$ as the parity symmetry is restored; this expectation is fulfilled, since $m_t \to 0$ in this limit.  At the other extreme, for large $M$, we expect to find $\delta g_{Lb}$ take on its SM value by having the factor within square brackets approach one.  This may be readily verified if we take the equivalent limit as $\mu \to \infty$ for fixed $m_t$:
\begin{equation}
\delta g_{Lb} (\mu \to \infty)  \to \frac{m_t^2}{16\pi^2 v^2}\left[1+\frac{1+\log(1/\mu^2)}{\mu^2}+{\cal O}(1/\mu^4)\right] \ ,
\label{eq:dglb}
\end{equation}
since in this limit $\sin\theta_L \to 1/\mu^2$, $\sin\theta_R \to 1/\mu$ and $m_T^2 / m_t^2 \to \mu^2$.  In other words, we find that adjusting the value of $M$ allows us to interpolate between the SM value for $\delta g_{Lb}$ at large $M$ and the absence of a radiative correction at small $M$.  While the limit of small $\mu$ is less useful, as we mentioned earlier, for completeness we note that 
\begin{equation}
\delta g_{Lb} (\mu \to 0)  \to \frac{m_t^2}{16\pi^2 v^2}\left[ \log(2/\mu) + \mu^2 (\frac34 + \frac12 \log(\mu/2))+{\cal O}(1/\mu^4)\right] \ ,
\label{eq:dglb2}
\end{equation}
since in this limit $\sin\theta_L \to (1/\sqrt{2})(1 - \mu^2/4)$, $\sin\theta_R \to (1 - \mu^2/8)$, and $m_T^2 / m_t^2 \to 4/\mu^2$.  This growth at small $\mu$ is visible in Fig.~(\ref{fig:gZbb}).  

We now use our results to compare the value of $g_{Lb}$ in our model (as a function of $\mu$ for fixed $m_t$) with the values given by experiment and the SM, as illustrated in Fig.~(\ref{fig:gZbb}).  The experimental \cite{:2005ema}  value $g_{Lb}^{ex}=-0.4182\pm0.0015$ corresponds to the thick horizontal line; the thin (red) horizontal lines bordering the shaded band show the $\pm 1 \sigma$ deviations from the experimental value.  We calculated the SM value using ZFITTER \cite{Bardin:1999yd,Arbuzov:2005ma} with a reference Higgs mass $m_h=115\text{ GeV}$, and obtain $g_{Lb}^{SM}=-0.42114$ (which matches the result in \cite{:2005ema}).  This is indicated by the dashed horizontal line, and may be seen to deviate from $g_{Lb}^{ex}$ by 1.96$\sigma$.   The (solid blue) curve shows how $g_{Lb}$ varies with $\mu$ in our model; we required $g_{Lb}$ to match the SM value with $m_t = 172$ GeV and $v = 246$ GeV as $\mu \to \infty$ and the shape of the curve reflects our results for $\delta g_{Lb}$ in Eq.~(\ref{eq:dglb}).   We see that $g_{Lb}$ in our model is slightly more negative than (i.e. slightly farther from the experimental value than) the SM value for $\mu > 1$, agrees with the SM value for $\mu = 1$, and comes within $\pm 1 \sigma$ of the experimental value only for $\mu < 1$.  Given the shortcomings of the small-$\mu$ limit, this is disappointing.

\subsection{Oblique Electroweak Parameters}

The flavor-universal corrections from new physics beyond the SM can be parametrized in a model independent way using the four oblique EW parameters $\alpha S,\ \alpha T,\ \alpha \delta,\ \Delta \rho$; the first two are the oblique
paramters  \cite{Peskin:1991sw,Altarelli:1990zd,Altarelli:1991fk} for models without additional electroweak gauge bosons, while the other two incorporate the effects of an extended electroweak sector. In general, the oblique parameters are related as follows \cite{Chivukula:2004af,Barbieri:2004qk} to the neutral-current 
\begin{align}
\label{NCamplitude}
-{\cal M}_{NC} = 4\pi \alpha \frac{Q Q'}{P^2} 
& +  \frac{(T^3-s_w^2 Q) (T'^3 - s^2 Q')}
	{\left(\frac{s_w^2c_w^2}{4\pi \alpha}-\frac{S}{16\pi}\right)P^2 +	\frac{1}{4 \sqrt{2} G_F}\left(1-\alpha T + 
	\frac{\alpha \delta}{4 s_w^2 c_w^2}\right)}  \\ 
& + \sqrt{2} G_F \, \frac{\alpha \delta}{s_w^2 c_w^2}\, T^3 T'^3 
+ 4 \sqrt{2} G_F  \left( \Delta \rho - \alpha T\right)(Q-T^3)(Q'-T'^3),\nonumber
\end{align} 
and charged-current electroweak scattering amplitudes 
\beq  
- {\cal M}_{\rm CC}
  =  \frac{(T^{+} T'^{-} + T^{-} T'^{+})/2}
             {\left(\frac{s_w^2}{4\pi \alpha}-\frac{S}{16\pi}\right)P^2
             +\frac{1}{4 \sqrt{2} G_F}\left(1+\frac{\alpha \delta}{4 s_w^2 c_w^2}\right)
            }
        + \sqrt{2} G_F\, \frac{\alpha  \delta}{s_w^2 c_w^2} \, \frac{(T^{+} T'^{-} + T^{-} T'^{+})}{2},
\label{CCamplitude}
\eeq
with $P^2$ a Euclidean momentum-squared.   In the DESM we may set $\Delta \rho = \alpha T$, because the model contains no extra $U(1)$ gauge group, and $\delta=0$, because there is no extra $SU(2)$ gauge group.  We therefore work purely in terms of $\alpha S$ and $\alpha T$ from here on.  We take the origin of the $\alpha S, \alpha T$ parameter space to correspond to the SM with $m_{H}=115\text{ GeV}$; this ensures that any non-zero prediction for the oblique parameters for a Higgs of this mass arises from physics beyond the SM.  At the one-loop level, the only new contributions to $\alpha S$ and $\alpha T$ in the DESM come from heavy fermion loops in the vacuum polarization diagrams indicated in Figure \ref{fig:Pi_ij}.   We therefore expect $\alpha S$ and $\alpha T$ to be of order a few percent\footnote{There are, in principle, additional oblique parameters such as $\alpha U$ that arise at higher order.  These will be suppressed relative to $\alpha S$ or $\alpha T$ by a factor of order $m_Z^2/m_{T}^2$; since we can see from Figure \ref{fig:masses} that $m_T > 2 m_t$, the suppression is by at least an order of magnitude and we shall neglect $\alpha U$ and its ilk from here on.} 

In this section, we will first separately derive expressions for $\alpha T$ and $\alpha S$ in DESM and see how each compares to current constraints from  \cite{Amsler:2008zzb}.  We then compare the DESM's joint prediction for $\alpha S$ and $\alpha T$ as a function of $\mu$ to the region of the $\alpha S - \alpha T$ plane that gives the best fit to existing data \cite{Amsler:2008zzb} and thereby derive a 95\% confidence level lower bound on $\mu$.  

\subsubsection{Parameter $\alpha T$}

\begin{figure}
\begin{center}
\includegraphics[width=2.5 in]{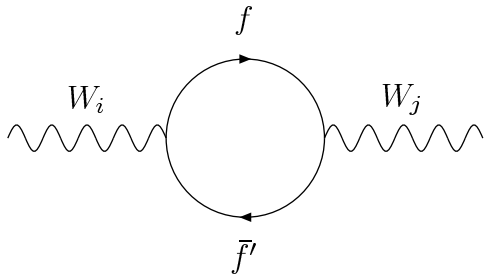}
\end{center}
\caption{Vacuum polarization diagram contributing to the oblique electroweak parameters. The indices $i,\ j=1,2,3,Q$, refer to weak (i = 1,2,3) or electromagnetic (Q) generators, while $f, f^\prime$ run over the appropriate combinations of $t$, $b$, $T$ and $\Omega$.}
\label{fig:Pi_ij}
\end{figure}

The custodial-symmetry-breaking parameter $\alpha T$ is defined as \cite{Peskin:1991sw}
\beq
\alpha T  = 
\left[ \frac{\Pi _{WW} \left( 0 \right)}{M^2_W} - \frac{\Pi _{ZZ} (0)}{M^2_Z}  \right],
\label{eq:Tdeff}
\eeq
where the contributions proportional to $g^{\mu\nu}$ in the vacuum polarization diagrams of Fig.~(\ref{fig:Pi_ij})
for the $W$ and $Z$ are labeled $\Pi_{WW}$ and $\Pi_{ZZ}$, respectively.  Each contribution sums
over various  $f \bar{f}^\prime$ pairs -- for $W$ we have $f \bar{f}^\prime =t\bar{b},\ T\bar{b},\ t \bar{\Omega},\ T \bar{\Omega}$; while for $Z$, we have $f \bar{f}^\prime=t\bar{t},\ T\bar{T},\ t\bar{T}, \Omega \bar{\Omega}, b\bar{b}, b \bar{\Omega}$.

 \begin{figure}[b]
\begin{center}
\includegraphics[width=4 in]{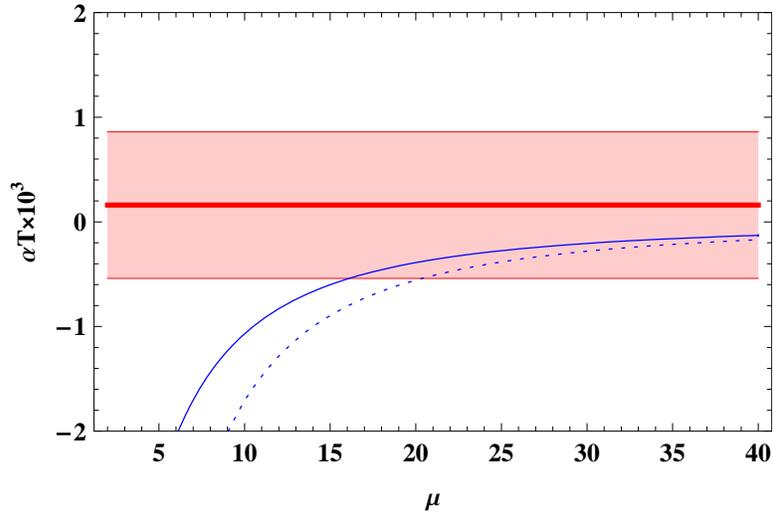} 
\end{center}
\caption{ The solid (blue) curve shows the DESM model's prediction for $\alpha T^{th}$ as a function of $\mu$. The horizontal lines show the optimal fit value of $\alpha T=0.16 \times 10^{-3}$ (thick solid line) and the relative $\pm 1\sigma$ deviations (solid lines bordering the shaded band) from \cite{Amsler:2008zzb} . The line $\alpha T = 0$ corresponds to the SM value (with $m_h=115$ GeV), by definition. The leading-log contribution to $\alpha T$, see Eq. (\ref{eq:leadingT}), is shown by the dotted curve.}
\label{fig:Drho}
\end{figure}

The analytical result for $\alpha T^{DESM}$ cannot be written in compact form;  the 
result\footnote{This is consistent with Eq. (33) in \cite{Pomarol:2008bh} when only the contributions from new fermions are included ($c_L = 0$ in the language of \cite{Pomarol:2008bh}).} in the limit $\mu>>1$ is:
\beq
\alpha T^{DESM}=\frac{3 m_t^2}{16\pi^2 v^2 }\left( {1 - 4\frac{{\ln \mu^2 }}{{\mu ^2 }} + \frac{{22}}{{ 3 \mu ^2 }}} \right)\,. 
\label{eq:Drho}
\eeq
One can see that, for $\mu\rightarrow \infty$, Eq.~(\ref{eq:Drho}) reproduces the leading SM result $\alpha T^{SM}(m_t)= 3 m_t^2 / (4\pi v)^2$ \cite{Peskin:1991sw}, as expected.   It interesting to note that the leading log contribution arising from the heavy states {\it reduces}\footnote{This does not violate the theorem \cite{Einhorn:1981cy, Cohen:1983fj} stating that $\Delta \rho \geq 0$ when mixing occurs only between particles of the same $T^3$ and hypercharge.  In the DESM, there is significant mixing between the $t^\prime_L$ and $T^\prime_L$ which have different $T^3$ and hypercharge values. As a result, we also expect significant GIM violation in the third generation.} the value of $\alpha T$.
This is to be expected, since the custodial symmetry is enhanced in the small-$\mu$ limit and
$\alpha T$ measures the {\it change} in the amount of isospin violation relative to the standard model. 

Subtracting the SM contribution from the top-quark, 
the numerical value of 
\begin{equation}
\alpha T^{th} = \alpha T^{DESM} - \alpha T^{SM}(m_t),
\end{equation}
as a function of $\mu$ is plotted as the solid blue curve in Fig.~(\ref{fig:Drho}); the dotted curve shows just the leading-log term (second term of Eq. (\ref{eq:Drho})).  
%Note that $\alpha T^{th} \to 0$ as $\mu \to \infty$ as we expect, since in that limit, the DESM reduces to the SM.  
The thick solid horizontal line corresponds to the best-fit value of $\alpha T = 0.16 \times 10^{-3}$ obtained by ref. \cite{Amsler:2008zzb} when setting $U = 0$; the two horizontal solid lines bordering the shaded band show the relative $\pm 1\sigma$ deviations from that central fit value.  Unlike the case of $\delta g_{Lb}$, the experimental constraints on $\alpha T$ clearly favor large values of $\mu$, closer to the SM limit.

By way of comparison, it is interesting to note that the authors of  \cite{Carena:2006bn,Carena:2007ua} studied the case where an SM-like weak-singlet top quark was in the same $SO(5)$ multiplet as extra quarks forming a weak doublet and concluded that this produced an experimentally-disfavored large negative contribution to $\alpha T$ at one loop.  Given that their  $SO(5)$ multiplet in 4D includes an $SO(4)=SU(2)_L\times SU(2)_R$ bi-doublet, our results are consistent with theirs
-- see Sec. (\ref{appxB}) below for further discussion.

\subsubsection{Parameter $\alpha S$}

\begin{figure}[b]
\begin{center}
\includegraphics[width=4 in]{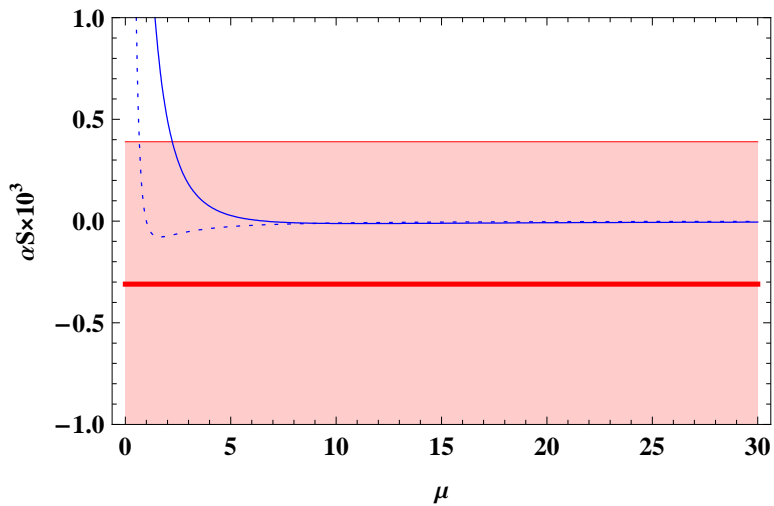}
\end{center}
\caption{The solid curve shows the DESM model's prediction for $\alpha S^{th}$ as a function of $\mu$. The horizontal lines show the optimal fit value of $\alpha S=-0.31 \times 10^{-3}$ (thick solid line) and the relative $\pm 1\sigma$ deviations (solid lines bordering the shaded band) from \cite{Amsler:2008zzb}. The line $\alpha S = 0$ corresponds to the SM value (with $m_h=115$ GeV), by definition. The leading-log contribution to $\alpha S$,
see Eq. (\ref{eq:leadingS}), is shown by the dotted curve.}
\label{fig:aS}
\end{figure}

The parameter $S$ is defined as \cite{Peskin:1991sw}
\beq
\alpha S = 16 \pi \alpha \left[ {\frac{d}{dq} \Pi _{33} \left( 0 \right) - \frac{d}{dq}\Pi _{3Q} \left( 0 \right)} \right],
\eeq
where $q$ is the gauge boson momentum. The complete expression for $\alpha S^{DESM}$ cannot be written in compact form; the limiting case where $\mu>>1$ is given by:
\beq
\alpha S^{DESM} = \frac{1}
{{6\pi }}\left( {3 + 2\ln \frac{{m_b }}
{{m_t }} + \frac{8}
{{\mu ^2 }}\left( {2 - \ln \mu } \right)} \right),
\label{eq:S}
\eeq
where we reintroduce a non-zero mass for the $b$ quark to cut off a divergence in the integral over the fermion loop momenta.  One can check that Eq.~(\ref{eq:S}) reproduces the SM result 
$\alpha S^{SM}(m_t,m_b)$ \cite{Peskin:1991sw} for $\mu \rightarrow \infty$.  Defining
\begin{equation}
\alpha S^{th} = \alpha S^{DESM}(\mu) - \alpha S^{SM}(m_t,m_b) \,,
\label{eq:delessth}
\end{equation}
we plot the result in Fig.~(\ref{fig:aS}), along with   the  value, $\alpha S=0.31\times 10^{-3}$, that provides an optimal fit to the data (for $U = 0$) and the $\pm 1\sigma$ relative deviations  \cite{Amsler:2008zzb}.  From Fig.~(\ref{fig:aS}) one can see that $\alpha S$ is within the $\pm 1\sigma$ bounds unless  $\mu < 3$; as with $\alpha T$, smaller values of $\mu$ are disfavored, though the constraint in this case is less severe.

\subsection{The $\alpha S$-- $\alpha T$ Plane}

\begin{figure}[bt]
	\begin{center}
\includegraphics[width=5 in]{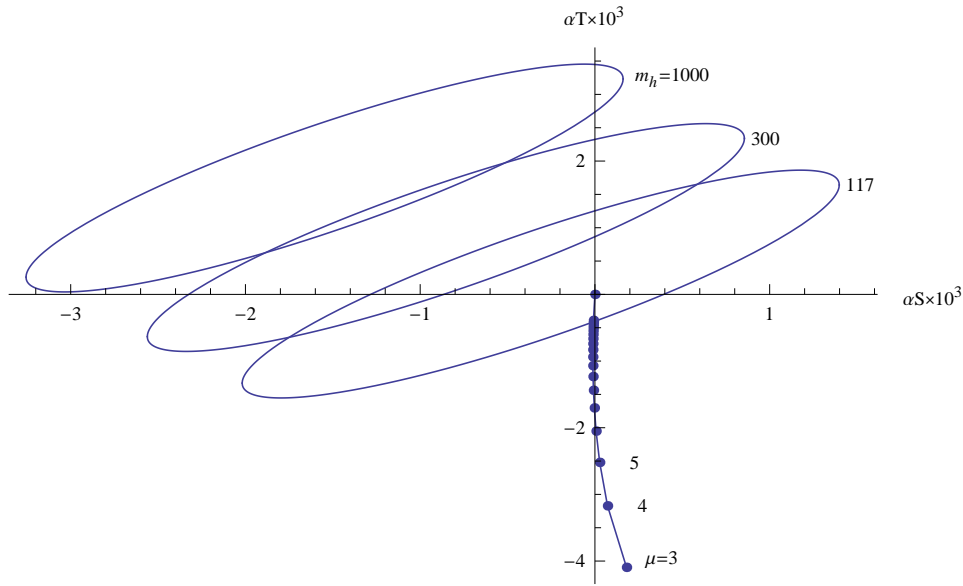}
\end{center}
\caption{The dots represent the theoretical predictions of the DESM  (with $m_h$ set to the reference value $115$ GeV), showing how the values of $\alpha S$ and $\alpha T$ change as $\mu$ successively takes on the values $3,\,4,\, 5, \,...,\,20,\,\infty$.   The three ellipses enclose the $95\%$CL regions of the $\alpha S$ - $\alpha T$ plane for the fit to the experimental data performed in \cite{Amsler:2008zzb}; they correspond to Higgs boson mass values of $m_h = 115\, {\rm GeV},\, 300\, {\rm GeV,\, and}\ 1\, {\rm TeV}$.   Comparing the theoretical curve with the ellipses shows that the minimum allowed value of $\mu$ is 20, for $m_h=115$ GeV. }
\label{fig:Ellipses}
\end{figure}

In Figure \ref{fig:Ellipses} we show the DESM predictions for  $[\alpha S^{th}(\mu), \alpha T^{th}(\mu)]$ from Eqs. (\ref{eq:delessth}, \ref{eq:Drho}) using $m_h = 115$ GeV,  and illustrating the successive mass-ratio values $\mu = 3,\, 4,\,...,20,\,\infty$; the point $\mu = \infty$ corresponds to the SM limit of the DESM and therefore lies at the origin of the $\alpha S$ - $\alpha T$ plane. 
On the same plane we also plot the elliptical curves that define the 95\% confidence level (CL) bounds on the $\alpha S$ - $\alpha T$ plane, relative to the optimal values of $\alpha S$ and $\alpha T$ found in \cite{Amsler:2008zzb}.
Reference  \cite{Amsler:2008zzb} provides the best-fit values and corresponding $\pm 1\sigma$ deviations for $m_h = 115$ GeV, $300$ GeV, along with the correlation matrix; we obtained the approximate values appropriate to $m_h = 1$ TeV by extrapolating based on the logarithmic dependence of $\alpha S$ and $\alpha T$ on $m_h$.  To calculate the 95\% CL ellipses, 
%we used the variables $a^i$, their best-fit values, their one-sigma deviations $\sigma^i$ and the correlation matrix $\rho^{ij}$ to define:
%
%\begin{equation}
%\chi^2 = \sum_{i,j} (a^i - a_{best-fit}^i) (\sigma^2)_{ij}^{-1} (a^j - a_{best-fit}^j),\ \ \ \ \ \ \ {\rm with\ \ } (\sigma^2)_{ij} \equiv \sigma_i \rho_{ij} \sigma_j\ ,
%\end{equation}
%
%and  
we solved the equation  $\Delta\chi^2 = \chi^2 - \chi^2_{min} = 5.99$, as appropriate to the $\chi^2$ probability distribution for two degrees of freedom. 

From this figure, we observe directly that the $95\%$CL lower limit on $\mu$ for $m_h=115\text{ GeV}$ is about 20, while for any larger value of $m_h$ the DESM with $\mu\leq 20$ is excluded at 95\%CL. In other words, the fact that a heavier $m_h$ tends to worsen the fit of even the SM ($\mu \to \infty$) to the electroweak data is exacerbated by the new physics contributions within the DESM.
The bound $\mu\geq 20$ corresponding to a DESM with a 115 GeV Higgs boson also implies, at 95\%CL, that $m_{T}\geq \mu \ m_t \cong 3.4\text{ TeV}$, so that the heavy partners of the top quark would likely be too heavy for detection at LHC.  

\section{Effective Field Theory}
\label{sec:Effective}

In this section, we use a simple effective field theory calculation to understand the size and form of
the non-SM corrections to $\alpha T$, $\alpha S$, and  the $Zb_L\bar{b}_L$ coupling
in the large-$\mu$ limit of the DESM.  
At large $\mu$, the fields $\Psi$ are approximately mass-eigenstates with mass
$m_T \approx M$. We proceed by using the equations of motion to 
``integrate out" the heavy $\Psi$ fields and construct the 
effective theory relevant for energies less than $M$ but greater than $m_t$.

\begin{figure}[bt]
	\begin{center}
\includegraphics[width=4.5 in]{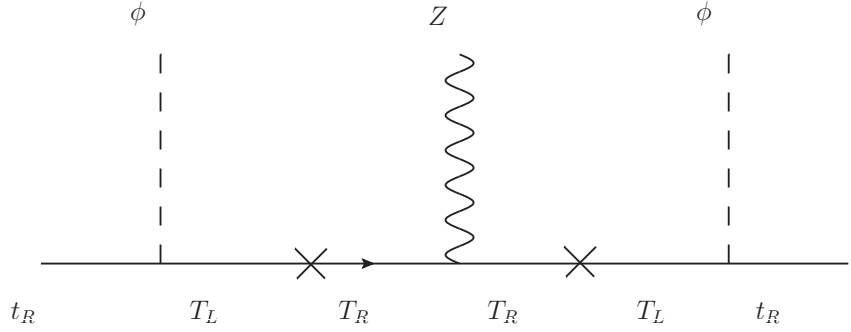}
\end{center}
\caption{\label{fig:tree} Neutral vacuum polarization diagrams giving the leading contribution to $\Delta\rho$ at scales below $M$, as a result of integrating out the new $\Psi$ fermions. The
non-standard vertex, identified by the black dot, arises from the operator in Eq.
(\protect\ref{eq:newterm}), and the crosses correspond to mass-insertions.}
\end{figure}

We start from the Lagrangian terms for the new fermion doublet $\Psi$:
\begin{equation}
{\cal L}_{\Psi} = i \bar{\Psi} D\!\!\!\!/ \,\Psi - M \bar{\Psi}\Psi - \lambda_t \text{Tr} \left( {\overline{\cal Q}_L} \cdot \Phi \right) t_R + h.c.
\label{eq:lag}
\end{equation}
In order to identify the terms involving $\Psi$, it is convenient to rewrite the Higgs field $\Phi$, introduced in Eq. (\ref{eq:higgsdef}), as
\begin{equation}
\Phi =  \left( \tilde{\phi}\,\, \phi \right)\, \ \ {\rm where}\,\,\, \phi \equiv \frac{1}{\sqrt{2}}\left( { \begin{array}{*{20}c} {i \sqrt{2}\, \phi^+} \\ {v + h - i \phi^0}\\ \end{array} } \right)\,\, ,\ \ {\rm and} \ \,  \tilde{\phi}= i\,\sigma_2\, \phi^\ast\, ,
\end{equation}
so that (ignoring terms not involving $\Psi$)
\begin{equation}
{\cal L}_{\Psi} = i \bar{\Psi} D\!\!\!\!/ \,\Psi - M \bar{\Psi}\Psi - \lambda_t \bar{\Psi}_L \phi\, t_R + h.c.
\label{eq:simplelag}
\end{equation}
Requiring the variation of ${\cal L}_{\Psi}$ with respect to $\bar{\Psi}_{L,R}$ to vanish yields the equations of motion
\begin{align}
i D\!\!\!\!/ \,\Psi_L - M \Psi_R & = \lambda_t \phi\, t_R\,,\\
i D\!\!\!\!/ \,\Psi_R - M \Psi_L & = 0~,
\end{align}
which we may solve iteratively in $1/M$. Doing so, we find
\begin{align}
\Psi_R & = -\,\frac{\lambda_t}{M}\phi\, t_R+{\cal O}\left(\frac{ (i D\!\!\!\! /\,)^2 \phi\, t_R}{M^3}\right)\\
\Psi_L & = - \frac{\lambda_t}{M^2} i D\!\!\!\!/ \,(\phi\, t_R)+{\cal O}\left(\frac{ (i D\!\!\!\! /\,)^3 \phi\, t_R}{M^4}\right)~.
\end{align}
Plugging these expressions into Eq. (\ref{eq:simplelag}), we obtain the 
non-SM terms in the low-energy effective theory
\begin{equation}
{\cal L}_{eff} = \frac{\lambda^2_t}{M^2} \bar{t}_R \phi^\dagger i D\!\!\!\!/ \,(\phi\, t_R) +  \ldots~,
\label{eq:newterm}
\end{equation}
where subsequent terms are suppressed by higher powers of $1/M^2$.
Note that, in terms of $\Phi$ (defined in Eq. (\ref{eq:higgsdef})), 
\begin{equation}
 \text{Tr}\left(\Phi^\dagger \partial^\mu \Phi\, \sigma^3\right)=2\left[(\partial^\mu \phi^\dagger) \phi - \phi^\dagger \partial^\mu \phi\right]~,
 \end{equation}
 and hence the operator in Eq. (\ref{eq:newterm}) violates custodial symmetry since it
 is the product of an $SU(2)_R$ singlet with one component of a triplet.\footnote{A similar computation shows that $\text{Tr}\left(\Phi^\dagger \partial^\mu \Phi\right) = 2\,\partial^\mu(\phi^\dagger \phi)$ is an $O(4)$ singlet.} Diagrammatically, this operator can be seen to
 arise from the process illustrated in Figure \ref{fig:tree}.

In unitary gauge this term gives rise to an ``anomalous" coupling
of the $Z$-boson to top-quarks, 
\begin{equation}
\frac{\lambda_t^2}{M^2} \bar{t}_R \phi^\dagger (i D\!\!\!\! /\,) \phi\, t_R\ \ \longrightarrow \ \ \frac{e \lambda_t^2 v^2}{4 s_w c_w  M^2}\  \bar{t}_R Z\!\!\!\! /\ \, t_R
\end{equation}
and is therefore capable of contributing to $\alpha T$ via oblique corrections to the $Z$ propagator.  Note that the sign
of this anomalous contribution is fixed by the gauge charge of the operator
$\phi\,t_R$, and that 
there is no induced correction to the $Wtb$ coupling and therefore no correction to the $W$ propagator.

\begin{figure}[bt]
	\begin{center}
\includegraphics[width=4.5 in]{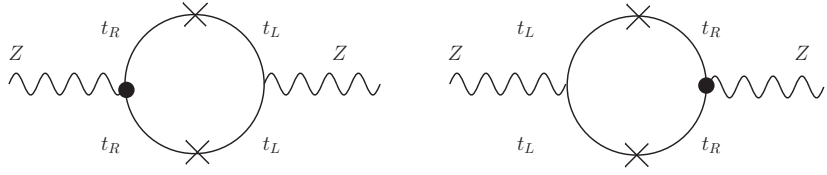}
\end{center}
\caption{\label{fig:W3} Neutral vacuum polarization diagram giving the leading contribution to $\Delta\rho$ at scales below $M$, as a result of integrating out the new $\Psi$ fermions. The
non-standard vertex, identified by the black dot, arises from the operator in Eqn.
(\protect\ref{eq:newterm}), and the crosses correspond to mass-insertions.}
\end{figure}

Given the definition of $\alpha T$ from Eq.  (\ref{eq:Tdeff}), it is clear that the leading non-SM contribution comes from the vacuum polarization diagrams shown in Figure \ref{fig:W3}, with one standard $Z\bar{t}t$ vertex plus one non-standard $Z\bar{t}t$ vertex due to integrating out the heavy fermions, $\Psi$. 
Evaluating this diagram yields
\begin{equation}
\Pi_{ZZ}(0) = 2 \cdot 3 \left[\frac{e^2 \lambda_t^2 v^2}{8 s_w^2 c^2_w M^2}\right] \left( - \frac{1}{8\pi^2}m_t^2 \ln \frac{m_t^2}{M^2}\right)
\label{eq:delmw3}
\end{equation}
where the 2 reflects the fact that either vertex could be the non-standard one, the 3 comes from summing over the colors of the internal quarks, and the factor in square brackets is the product of the vertex coefficients. The factor in parentheses is the result of the loop integral, with the large log arising from the separation between the scales of the $\Psi$ and top masses.   Since there is no correction to $\Pi_{WW}(0)$ from integrating out the heavy fermions $\Psi$, we conclude that 
\begin{equation}
\alpha T^{eff} =  - \frac{\Pi_{ZZ}(0)}{M^2_Z} = \frac{3 m_t^2}{16\pi^2 v^2} \left[ -4 \frac{m_t^2}{M^2} \ln \frac{M^2}{m_t^2} \right]
\label{eq:leadingT}
\end{equation}
where we have used $m_t = \lambda_t v / \sqrt{2}$, as appropriate to the large-$M$ limit of the DESM.  Recalling that $\mu \approx M / m_t$ for large $\mu$, we see that our effective theory result is identical to the leading (large log) correction to $\alpha T^{DESM}$ from non-SM physics obtained earlier in Eq. (\ref{eq:Drho}). 

\begin{figure}[bt]
	\begin{center}
\includegraphics[width=2.0 in]{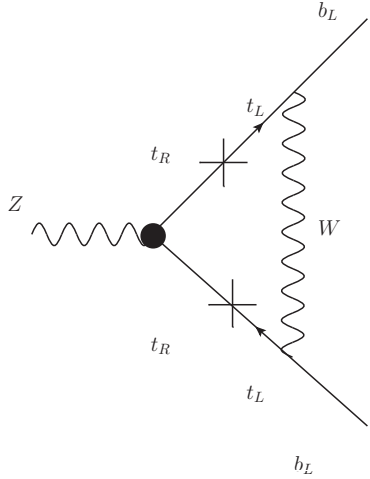}
\end{center}
\caption{\label{fig:zbb-eff} Leading-log non-standard contribution to the $Zb\bar{b}$
vertex in the low-energy effective theory for scales below $M$ yet above $m_t$. The
non-standard vertex, identified by the black dot, arises from the operator in Eqn.
(\protect\ref{eq:newterm}), and the crosses correspond to mass-insertions.}
\end{figure}

Similarly, calculation of $\Pi_{33}$ and $\Pi_{3Q}$ in the effective
theory yields,
\begin{equation}
\alpha S^{eff} = \frac{1}{6\pi}\left[-\,\frac{8 m^2_t}{M^2} \log\frac{M^2}{m^2_t}\right]~,
\label{eq:leadingS}
\end{equation}
in agreement with Eq. (\ref{eq:S}).
The leading-log contribution to the $Z\to b\bar{b}$ vertex can analogously be understood
as arising from the diagram illustrated in Figure \ref{fig:zbb-eff}, yielding
\begin{equation}
\delta g^{eff}_{Lb} = \frac{m_t^2}{16\pi^2 v^2}\left[
\frac{m_t^2}{M^2}\log\frac{m_t^2}{M^2}\right]~\, ,
\label{eq:zbbleading}
\end{equation}
in agreement with the leading $\log$ of Eq.~(\ref{eq:dglb}).

Since all of the effects on all three observables ($\alpha T$, $\alpha S$ and $\delta g_{Lb}$)
arise from the operator in eqn. (\ref{eq:newterm}) in the DESM, we see why
the size and signs of these effects are correlated as observed in the previous sections.

\section{Relation to Extra-Dimensional Models}
\label{appxB}

In extra-dimensional models  there are Kaluza-Klein (KK) excitations for the gauge fields and the fermion fields.  In
the top-quark mass generating sector, these extra fields give rise to additional contributions
to  $\alpha S$, $\alpha T$, and to the $Zb_L\bar{b}_L$ coupling \cite{Agashe:2003zs}. Incorporating an approximate 
$O(4)\times U(1)_X$ symmetry has been proposed as a mechanism
to control the size of these additional contributions to $\alpha T$ and the $Zb_L\bar{b}_L$ coupling \cite{Agashe:2006at}.
Such effects have been thoroughly analyzed in the case of an $SO(5)$  gauge-Higgs model in ref.~\cite{Carena:2006bn,Carena:2007ua}; here we review the top-quark Yukawa sector of that extra-dimensional model and compare it with our model.  We also comment on the model-independent analysis in ref.~\cite{Pomarol:2008bh} that discusses these effects.

In a 5D model of the kind discussed in \cite{Carena:2006bn,Carena:2007ua}, the custodial symmetry is
imposed on the bulk interactions. Crucially, however, the standard $SU(2)_L$ quark doublet $q_L$ and the new doublet $\Psi_L$, which make up the (2,$2^*$) multiplet ${\cal Q}_L$, have different boundary conditions: $(+,+)$ for $q_L$ and $(-+)$ for $\Psi_L$. These allow a $q_L$ zero mode to exist, but forbid a $\Psi_L$ zero mode. The singlet quark state $t_R$ has boundary conditions $(+,+)$ and thus also has a zero mode. As a consequence, in the unbroken electroweak phase the 4D fermions consist of a massless $SU(2)_L$ doublet $q_{0L}$, a massless singlet $t_{0R}$, and a tower
of  KK Dirac fermions which we will denote by  $q_n$, $t_n$, and $\Psi_n$ for $n=1,2,\ldots$. (Recall, again,
that the $\Psi$ has no zero mode, due to the boundary conditions.) 
By incorporating the $O(4) \times U(1)_X$ symmetry in the bulk but breaking it via boundary
conditions, the spectrum and top-quark mass generating interactions are essentially the same as those
in the standard model for the zero modes, but (especially in the limit of large $O(4)$-symmetric bulk mass)
are approximately custodially symmetric for the
KK modes.  

As we now demonstrate, the enhanced custodial symmetry of the KK sector mitigates the size of additional
contributions to $\alpha T$ and the $Zb_L\bar{b}_L$ coupling, while the zero-mode contributions approximately
give rise to the usual standard model contributions.
Consider the 4D Lagrangian for the top-quark mass generating 
sector of zero-modes and the first level of KK fermions, which has the form
\begin{eqnarray}
{\cal L}_{4D} &=& \bar{q}_{0L}\, i\, \slashed{D} q_{0L} + \bar{t}_{0R}\, i\, \slashed{D} t_{0R} + \bar{q}_1\, i\, \slashed{D} q_1 
+ \bar{\Psi}_1\, i\, \slashed{D} \Psi_1 + \bar{t}_1\, i\, \slashed{D} t_1
-M_{\Psi_1}\, \bar{\Psi}_1\, \Psi_1 - M_{q_1}\, \bar{q}_1\, q_1 - M_{t_1}\, \bar{t}_1\, t_1 \nonumber \\
&-& \Big(\lambda_{00}\, \bar{q}_{0L}\, \tilde{\phi}\, t_{0R} + \eta_{10}\, \bar{\Psi}_{1L}\, \phi\, t_{0R}
+ \lambda_{01}\, \bar{q}_{0L}\, \tilde{\phi}\, t_{1R} + \eta_{11}\, \bar{\Psi}_{1L}\, \phi\, t_{1R}
-\lambda_{10}\, \bar{q}_{1L}\, \tilde{\phi}\, t_{0R}-\lambda_{11}\, \bar{q}_{1L}\, \tilde{\phi}\, t_{1R} \nonumber \\
&-&\lambda^\prime_{11}\, \bar{q}_{1R}\, \tilde{\phi}\, t_{1L}-\eta^\prime_{11}\, \bar{\Psi}_{1R}\, \phi\, t_{1L} + {\rm h.c.}\Big) \ ,
\label{eqn:extrad-fermions}
\end{eqnarray}
where the masses $M_{q_1}$, $M_{\Psi_1}$, and $M_{t_1}$ have approximately the same size (of order a TeV). The Yukawa couplings depend on the Higgs and fermion profiles, and therefore generally differ from one another, although they are all expected to be of the order of the SM top-Yukawa coupling. Notice that the $O(4)$ symmetry in the bulk, which relates the same-level KK modes of $q_L$ and $\Psi_L$,  implies
\begin{eqnarray}
\lambda_{1i}\simeq \eta_{1i};
\label{eq:symm1}
\end{eqnarray}
thus, for large values of the common (extra-dimensional) bulk mass, we find 
\begin{equation}
M_{q_1}\simeq M_{\Psi_1}~. 
\label{eq:symm2}
\end{equation}
This contrasts with the situation in our model where  $\lambda_{00}=\eta_{10}$ because the $O(4)$ symmetry relates $q_{0L}$ and $\Psi_{1L}$; in the 5D models, $q_{0L}$ and $\Psi_{1L}$ belong to different KK levels, and such a symmetry does not arise.

Integrating out the heavy fermions ($q_1$, $\Psi_1$, and $t_1$)
at tree-level gives rise to the higher-dimensional operators
\begin{eqnarray}
{\cal L}_{\rm eff} = \frac{\lambda_{10}^2}{M_{q_1}^2}\bar{t}_{0R}\, \tilde{\phi}^\dagger \, i\, \slashed{D}\left(\tilde{\phi}\, t_{0R}\right)
+ \frac{\eta_{10}^2}{M_{\Psi_1}^2}\bar{t}_{0R}\, \phi^\dagger \, i\, \slashed{D}\left(\phi\, t_{0R}\right)
={\rm Tr} \left[
\begin{pmatrix}
\frac{\lambda^2_{10}}{M^2_{q1}} & 0\\
0 & \frac{\eta^2_{10}}{M^2_{\Psi 1}}
\end{pmatrix}
\bar{t}_{0R}\, \Phi^\dagger \, i\, \slashed{D}\left(\Phi\, t_{0R}\right)
\right]
 \ .
\label{eq:eff5d}
\end{eqnarray}
The operator proportional to $\eta^2_{10}$ is the same as in Eq.~(\ref{eq:newterm}), and diagrammatically comes from the exchange of a $\Psi_1$. The operator proportional to $\lambda^2_{10}$ is not present in our model, and comes from $q_1$ exchange. The exchange of a heavy singlet $t_1$ does not induce any new operator at tree-level, and thus plays no relevant role at low energy. Therefore, the only new ingredient in the 5D model, relative to ours, is the presence of the heavy doublet $q_1$. After electroweak symmetry breaking,
both of these operators contribute to the $Zt_R\bar{t}_R$ coupling, which yield (at one-loop)
 corrections to $\alpha S$, $\alpha T$, and the $Zb_L \bar{b}_L$ coupling as we demonstrated
 in Sec. (\ref{sec:Effective}). 

Notice, in particular,  that the $\bar{t}_{0R} Z\!\!\!\! /\ \, t_{0R}$ vertex receives contributions of opposite sign from the two operators in  Eq.~(\ref{eq:eff5d}), due to the different $T_3$ charges of $\langle \phi\rangle$ and 
$\langle \tilde{\phi}\rangle$. Furthermore, Eqs.~(\ref{eq:symm1})-(\ref{eq:symm2}) tell us that these contribution are approximately equal in absolute value, so they will tend to cancel each other. As displayed in the last equality in Eq. (\ref{eq:eff5d}), in the custodially-symmetric limit,
 $\lambda^2_{10}/M^2_{q1} = \eta^2_{10}/M^2_{\Psi 1}$ and the combination of operators is manifestly $O(4)$ symmetric. 
 
 Thus we have the following picture. In the 5D model the $q_1$ and $\Psi_1$ KK modes approximately give equal but opposite contributions to $Z\bar{b}_L b_L$. This is true because the fermion KK modes form approximate (2,$2^*$) multiplets under $O(4)$. 
 The remaining and dominant contribution is the SM one. Because of the cancellation of the KK contributions to the $Z b_L \bar{b}_L$ coupling in these 5D models, it is possible to find a region of parameter space in which the KK fermion contributions to $\alpha S$, $\alpha T$, and the $Zb_L\bar{b}_L$ coupling,
is small. In this case, the extra fermions in the 5D model can be relatively light \cite{Carena:2006bn,Carena:2007ua}.

By contrast, in the DESM  it is $q_L$ and $\Psi_L$ which belong to the same $O(4)$ multiplet, and
the contributions from $\Psi$ and the standard model top-bottom loops
tend to cancel in their contributions to electroweak processes. As shown above, however,
the custodial violation characteristic of the standard model is required by electroweak data, and we are therefore
pushed into the regime where the extra fermions of the DESM are very heavy.

Ref.~\cite{Pomarol:2008bh} also investigated the low-energy effective theories that arise when integrating out the new vector and fermion states generically present in extra-dimensional or strongly-coupled models of electroweak symmetry breaking that feature the custodial symmetries protecting $Zb\bar{b}$.  Their results for the effects of integrating out the new fermions in this framework are consistent with the analysis given above, confirming the broad applicability of this aspect of our findings.  Specifically, they also find that integrating out the new fermions gives rise to the operator in Eq. (\ref{eq:newterm}) and that this operator affects $\alpha T$ as shown in Eq. (\ref{eq:Drho}).  Our more stringent lower bound on $m_T$ arises from the different central values and tighter $S-T$ correlations in the 2008 data \cite{:2005ema,Amsler:2008zzb}, compared with the 2004 data they employed \cite{Barbieri:2004qk}.  In the gauge sector, ref.~\cite{Pomarol:2008bh} finds that additional custodial-symmetry-violating operators arise from integrating out the extra vector bosons, provided that custodial-symmetry violation is present in the boundary conditions.  These extra operators contribute to $g_{Lb}$, $\alpha T$ and $\alpha S$, and can also be adjusted to achieve agreement with experimental bounds.

\section{Conclusions}
\label{sec:Conclusion5}

We have introduced the doublet-extended standard model (DESM) as a simple realization of the idea \cite{Agashe:2006at} of using custodial symmetry to protect the $Zb_L\bar{b}_L$ coupling ($g_{Lb}$) from receiving large radiative corrections.  In this toy model, all terms of dimension-4 in the top-quark mass generating sector
obey a global $O(4)\times U(1)_X$ symmetry, which includes a parity symmetry protecting $g_{Lb}$ from radiative corrections.  That global symmetry is softly broken to its $SU(2)_L \times U(1)_Y$ subgroup by a Dirac mass term for the extra fermion doublet that incorporates the heavy partner of the top quark.  Varying the size of this Dirac mass $M$ allows the model to interpolate between the $O(4)\times U(1)_X$-symmetric case ($M=0$) in which $\delta g_{Lb} = 0$ and the SM-like case ($M \to \infty$) in which the one-loop corrections to $g_{Lb}$ are as in the SM, and
enabled us to investigate the degree to which
the custodial symmetry of the top-quark mass generating sector can be enhanced.  
By comparing the predictions of the DESM with experimental constraints on the oblique parameters $\alpha S$ and $\alpha T$ from \cite{Amsler:2008zzb}, we found the DESM to be consistent with experiment only for $\mu>20$ at 95\%CL, with a Higgs mass $m_h=115$ GeV. The bound on $\mu$ translates into the 95\% CL lower bound of $3.4\text{ TeV}$ on the masses of the extra quarks -- placing them out of reach of the LHC. This result demonstrates 
that electroweak data strongly limits the
amount by which the custodial symmetry of the top-quark mass generating sector can be enhanced relative
to the standard model. 

In addition, we performed an effective field theory analysis of the energy regime between $M$ and $m_t$. We 
demonstrated that the leading contributions to $\alpha S$,  $\alpha T$ and the $Zb_L \bar{b}_L$ coupling in the DESM arise from an effective operator coupling right-handed top-quarks to the $Z$-boson.  
Extending our effective field theory analysis, we considered extra-dimensional models in which
the enhanced custodial symmetry is invoked to control the size of additional contributions to $\alpha T$ and
the $Zb_L\bar{b}_L$ coupling, while leaving the standard model contributions essentially unchanged
 \cite{Carena:2006bn,Carena:2007ua}.  In such models, the global $O(4)$ symmetry causes a cancellation among contributions to the effective operator coupling $t_R$ to the $Z$, allowing relatively light 5D fermions to be consistent with experiment. Finally, our results are also consistent with those of ref.~\cite{Pomarol:2008bh}
which confirms that the toy DEWSB model illustrates the electroweak physics operative in a broad class of models.

\begin{acknowledgments}
This work was supported in part by the US National Science Foundation under grant PHY-0354226.  RSC and EHS acknowledge the support of the Aspen Center for Physics where part of this work was completed. We also
thank the anonymous referee for suggestions which prompted us to clarify our presentation and conclusions.
\end{acknowledgments}

\end{document}